\documentclass[aps,superscriptaddress,notitlepage,10pt,a4paper]{revtex4-1}

\usepackage[pdftex]{graphicx}
\usepackage{amsmath}
\usepackage{hyperref}
\usepackage{times}
\usepackage[centering,hmargin=2cm,vmargin=2.5cm]{geometry}

\hypersetup
{%
pdftitle = {Scattering of an exponential pulse by a single atom},
pdfauthor = {Markus Sondermann},
colorlinks = {false}
}

\begin{document}

\title{Scattering of an exponential pulse by a single atom}

\author{Markus Sondermann}
\email{Markus.Sondermann@physik.uni-erlangen.de}
\author{Gerd Leuchs}
\affiliation{Institute of Optics, Information and Photonics,
  University of Erlangen-Nuremberg, 91058 Erlangen, Germany}
\affiliation{Max Planck Institute for the Science of Light, 91058
  Erlangen, Germany}

\begin{abstract}
We discuss the scattering of a light pulse by a single atom in
free space using a purely semi-classical framework.
The atom is treated as a linear elastic scatterer allowing to 
treat each spectral component of the incident pulse separately.
For an increasing exponential pulse with a dipole radiation pattern
incident from full solid angle the spectrum resulting from
interference of incident and scattered components is a decreasing
exponential pulse.
\end{abstract}

\maketitle

\section{Introduction}

Recent experiments devoted to the coupling of single two-level
emitters to the light field in free space have been mainly concerned
with the scattering of monochromatic coherent laser 
beams~\cite{vamivakas2007,wrigge2008,tey2008-np,aljunid2009,
slodicka2010,pototschnig2011}.
In these experiments the incident light field has been so weak that
the steady-state population of the emitter's excited state was
negligible.

Unlike in the above cases, experiments with non-monochromatic incident
light fields have been performed in
Refs.~\cite{rezus2012,aljunid2013}.
In Ref.~\cite{rezus2012} the incident light field was constituted by a
constant stream of single photons from a source molecule, i.e. pulses
with an exponentially decaying temporal envelope having a spectral width
matching the one of the transition of the target molecule.
These pulses are expected to create a non-negligible excited state
population~\cite{stobinska2009}. 
However, this quantity has not been measured in Ref.~\cite{rezus2012}.
Instead, the extinction of the photon stream has been monitored. 
In Ref.~\cite{aljunid2013} the incident radiation was a coherent state
light pulse with an increasing exponential envelope and a finite
amount of atomic  excitation was measured.

Here, we want to establish a link between elastic scattering
experiments -- usually prohibitive of atomic excitation -- and the
absorption of single photons or weak coherent state pulses. 
To do so, we treat the atom as a driven harmonic oscillator with a
driving force that is weak enough to keep the oscillator's response in
the linear regime. 
This approach is motivated by the close analogy between a coherently
driven classical harmonic oscillator and a single atom driven by a
single photon~\cite{heugel2010}.
We will decompose a light pulse into its spectral
components.
Each of these components constitutes a monochromatic wave.
It is assumed that the scattering of each of these waves is
completely elastic, enabling interference with the corresponding
incident spectral component.
The resulting spectrum then determines the temporal response of the
atom.

Of course, this treatment is not applicable to cases where more than a
single photon is contained in the incident pulse. 
As is evident from fully quantum mechanical treatments
\cite{wang2011}, the upper state population and hence the
electromagnetic field will exhibit Rabi oscillations.
This is clearly not covered by the treatment discussed in this paper.
However, even a pulse containing the energy of a single photon
resonant with the atomic transition induces a non-negligible amount of
excited state population.
For such a pulse with an effective length of the excited state
lifetime, the Rabi frequency can be as large as twice the spontaneous
emission rate~\cite{leuchs2013o}.
This corresponds to a saturation parameter of $S=8$, i.e. an excited
state population of $\rho_\textrm{ee}=4/9$.
This finding suggests that the calculations presented below are only
meaningful for pulses containing much less than a single photon,
e.g. strongly attenuated coherent states as prepared in
Refs.~\cite{golla2012,aljunid2013}.

However, the value of $\rho_\textrm{ee}$ obtained from $S$ is a steady
state quantity. 
On the time scale of the excited state lifetime -- and single photon
pulses used in free space experiments are typically of this duration
-- the steady state is not yet reached.
Rather the excited state population has yet to build up from zero.
Therefore, one could expect that the fully elastic treatment is
justified during almost the complete duration of the pulse, especially
if the amplitude of the incident pulse itself increases slowly. 
This is the case for exponentially increasing pulses, which have been
predicted to excite an atom with full
efficiency~\cite{sondermann2007,stobinska2009,wang2011}.

The paper is organized as follows: In the next section we will briefly
revisit the scattering of a monochromatic wave.
Then in Sec.~\ref{sec:scatexp} this framework will be applied to all
spectral components of an exponentially increasing pulse, yielding the
spectrum of the temporal atomic response.
The paper will close with a brief discussion.

\section{Scattering by an atomic dipole}
\label{sec:scatmono}

At first, the scattering of a monochromatic wave by an atomic dipole
is reviewed. 
The atom is taken as a two-level system and considered to be in the
steady state under the monochromatic driving field.
The derivation of the respective formulas is given in
Ref.~\cite{sondermann2013p}.  Here we just recall the results
relevant to this paper.

The power scattered by the atom is given by
\begin{equation}
\label{eq:Psc}
P_\mathrm{sc}= \frac{4P\cdot\Omega\eta^2}
{(4\Delta^2/\Gamma^2+1)(1+s)^2}
 \quad . 
\end{equation}
$P$ is the power of the incident beam.
$\Omega$ is the solid angle of the focused field weighted by the angular
intensity pattern $I(\vartheta,\phi)$ of the atomic transition dipole
moment~\cite{sondermann2008}:
\begin{equation}
\Omega=\frac{\int_{\phi_\textrm{foc}}\int_{\vartheta_\textrm{foc}}
I(\vartheta,\phi)\sin\vartheta\, d\vartheta d\phi}{8\pi/3}
\, .
\end{equation}
It is given as a normalized quantity ($0\le\Omega\le1$) with the case
$\Omega=1$ corresponding to focusing from full solid angle.
$\eta$ is the spatial mode overlap of the incident field with the
field emitted by the atomic dipole.
The overlap is integrated over and normalized to only the part of the
solid angle covered by the incident beam.
$\Delta=\omega-\omega_0$ is the detuning from the atomic resonance
$\omega_0$, and $\Gamma$ is the spontaneous emission rate.
Finally, $s$ is the saturation parameter which depends on all of
the other parameters given above, see Ref.~\cite{sondermann2013p}.
However, since here we are interested in the regime of elastic
scattering we set $s=0$ and have
\begin{equation}
\label{eq:Psc0}
P_\mathrm{sc}= \frac{4P\cdot\Omega\eta^2}
{4\Delta^2/\Gamma^2+1}
 \quad . 
\end{equation}
This equation is equivalent to the findings of
Refs.~\cite{zumofen2008, tey2009, zumofen2009}, once one identifies the
quantity $4\Omega\eta^2$ with the scattering ratio used in these
papers.

$P_\mathrm{sc}$ can be written as $P_\mathrm{sc}=
\mathrm{const.}\times |E_\mathrm{sc}|^2$, where
$E_\mathrm{sc}=A_\mathrm{sc}\cdot e^{i\varphi_\mathrm{sc}}$ is the
  complex amplitude of the dipole wave scattered by the atom.
Neglecting proportionality constants, we can write
$A_\mathrm{sc}=\sqrt{P_\mathrm{sc}}$, i.e. 
\begin{equation}
\label{eq:Asc0}
A_\mathrm{sc}= \frac{2A\cdot\sqrt{\Omega}\eta}
{\sqrt{4\Delta^2/\Gamma^2+1}}
 \quad ,
\end{equation}
where $A=\sqrt{P}$ is the modulus of the incident field amplitude.
$\varphi_\mathrm{sc}$ is the phase of the the scattered wave
  \emph{relative} to the phase of the incident field.
It is given by~\cite{sondermann2013p,zumofen2008}
\begin{equation}
\varphi_\textrm{sc}=\arctan\left(\frac{2\Delta}{\Gamma}\right)
+\frac{\pi}{2}\, ,
\end{equation}
as has been confirmed in a recent experiment~\cite{jechow2013}.
With $A_\mathrm{sc}$ and $\varphi_\textrm{sc}$ we have all quantities
at hand that are needed to calculate the field resulting from the
superposition of incident field and scattered field.

In almost all of the recent experiments dealing with light-matter
interaction in free space, the light scattered by the atom and the
incident radiation are collected by optics spanning nominally the same
solid angle fraction as the device focusing the incident radiation.
The part of the solid angle not covered by the collection optics
is governed by the scattered field alone.
The power emitted into this part of the solid angle is
\begin{equation}
\label{eq:Pcompl}
P_{\mathrm{sc},\overline\Omega}= (1-\Omega)\cdot
\frac{\Gamma^2\cdot\Omega\eta^2} 
{\Delta^2+\Gamma^2/4}\cdot A^2
\end{equation}
with the corresponding complex field amplitude 
\begin{equation}
\label{eq:Ecompl}
E_{\mathrm{sc},\overline\Omega}= \frac{\Gamma\eta\sqrt{\Omega(1-\Omega)}}
{\sqrt{\Delta^2+\Gamma^2/4}}\cdot Ae^{i\varphi_\mathrm{sc}} \quad .
\end{equation}

The fraction of the scattered power emitted towards the collection
optics is 
\begin{equation}
\label{eq:Psc_omega}
P_{\mathrm{sc},\Omega}= 
\frac{\Gamma^2\Omega^2\eta^2}{\Delta^2+\Gamma^2/4}\cdot A^2
\quad .
\end{equation}
In this solid angle fraction, where the scattered light interferes with
the rediverging incident light, an additional $\pi/2$ shift related to the Gouy
phase has to be
considered~\cite{tyc2012,zumofen2008,pototschnig2011,aljunid2009}. 
We do this by writing 
\begin{equation}
\varphi_\textrm{sc}=\arctan\left(\frac{2\Delta}{\Gamma}\right)
+\pi \quad .
\end{equation}
The corresponding field amplitude is 
\begin{equation}
\label{eq:Esc_omega}
E_{\mathrm{sc},\Omega}= \frac{\Gamma\eta\Omega}
{\sqrt{\Delta^2+\Gamma^2/4}}\cdot Ae^{i\varphi_\mathrm{sc}} \quad .
\end{equation}
We assume that all of the incident radiation is collected as well.
However, only a part of the scattered radiation can interfere
with the incident field.
The corresponding power fraction is proportional to $\eta^2$.
Thus, the field component due to interference is
\begin{equation}
\label{eq:Eomega_coh}
E_{\Omega,\mathrm{coh}}=\frac{\Gamma\eta^2\Omega}
{\sqrt{\Delta^2+\Gamma^2/4}}\cdot Ae^{i(\varphi_\mathrm{sc}+\varphi_0)} 
+ Ae^{i\varphi_0} 
\quad ,
\end{equation}
where we have also allowed for some arbitrary relative phase
$\varphi_0$ of the incident field.
The corresponding power reads
\begin{equation}
\label{eq:Pomega_coh}
P_{\Omega,\mathrm{coh}}=
\left[ 1 + \frac{\Gamma^2}{\Delta^2+\Gamma^2/4}\left(
\Omega^2\eta^4-\Omega\eta^2 \right)\right]
\cdot A^2 
\quad . 
\end{equation}

\begin{figure}
\centerline{\includegraphics{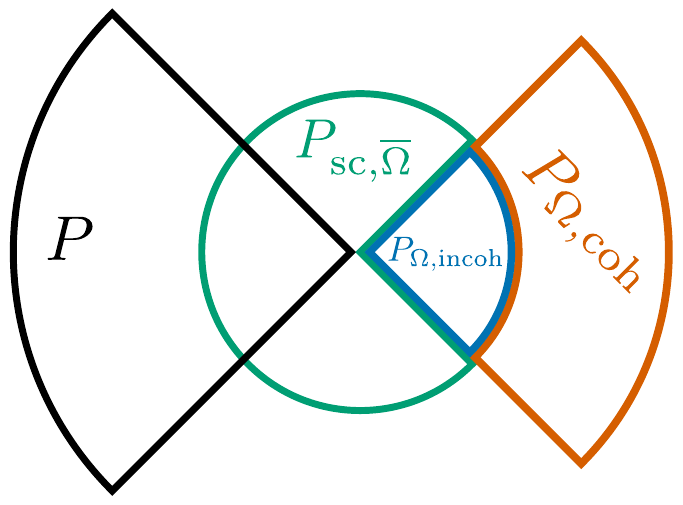}}
\caption{\label{fig:powers}
Illustration of the different power fractions involved in elastic
scattering by a single atom.
The incident field is focused from a solid angle fraction $\Omega$
with power $P$.
In transmission, the incident field interferes with the scattered
field, with the solid angle fraction on which the interference occurs
again being $\Omega$.
Since the spatial mode overlap between scattered and incident field is
in general not perfect, one has to account for a power fraction
$P_{\Omega,\mathrm{coh}}$ due to this interference and a remaining
fraction of the scattered light with power
$P_{\Omega,\mathrm{incoh}}$.
The light scattered into part of the solid angle complementary to the
transmission one is of power $P_{\textrm{sc},\overline\Omega}$.
}
\end{figure}

For completeness, we also give the respective expressions for the
fraction that does not interfere with the incident field.
The complex field amplitude reads
\begin{equation}
\label{eq:Eomega_incoh}
E_{\Omega,\mathrm{incoh}}=\frac{\Gamma\Omega\eta\sqrt{1-\eta^2}}
{\sqrt{\Delta^2+\Gamma^2/4}}\cdot Ae^{i\varphi_\mathrm{sc}} 
\end{equation}
with the corresponding power
\begin{equation}
\label{eq:Pomega_incoh}
P_{\Omega,\mathrm{incoh}}=
\frac{\Gamma^2\Omega^2\eta^2(1-\eta^2)}{\Delta^2+\Gamma^2/4}\cdot
A^2 
\quad . 
\end{equation}
With some algebra it is easy to check that energy is conserved.
The meaning of the different power fractions is illustrated in
Fig.~\ref{fig:powers}.

\section{Incident pulse with increasing exponential envelop}
\label{sec:scatexp}

In the following we treat the case of an incident wave with carrier
frequency $\omega_0$ and an exponentially 
increasing intensity envelop. 
The time constant of the exponential shall be the life time of the upper
atomic state and the wave ends at $t=0$.
The time dependent field amplitude of such a wave is given by
$A_0\cdot\exp(\Gamma t/2)\cdot H(-t)$ with $H(t)$ being the step
function.
The spectrum of this wave is given by 
\begin{equation}
S(\Delta)= A_0 \cdot \frac{1}{\Gamma/2 + i\Delta}
\end{equation}
with the spectral amplitude $A(\Delta)=A_0/\sqrt{\Delta^2+\Gamma^2/4}$
and the relative spectral phase
$\varphi_0(\Delta)=\arctan(-2\Delta/\Gamma)$.

For this incident pulse spectrum the resulting spectral field
amplitudes can be written as 
\begin{equation}
\label{eq:Ecompl_exp}
E_{\mathrm{sc},\overline\Omega}(\Delta)= A_0 \cdot
i\frac{\eta\Gamma\sqrt{\Omega(1-\Omega)}}
{\Delta^2+\Gamma^2/4} \quad ,
\end{equation}
\begin{equation}
\label{eq:Einc_exp}
E_{\Omega,\mathrm{incoh}}(\Delta)= A_0 \cdot
i \frac{\Gamma\Omega\eta\sqrt{1-\eta^2}}
{\Delta^2+\Gamma^2/4} \quad ,
\end{equation}
\begin{equation}
E_{\Omega,\mathrm{coh}}(\Delta)=A_0\cdot 
\frac{\Gamma(1/2-\Omega\eta^2) -i\Delta}{\Delta^2+\Gamma^2/4}
\quad .
\end{equation}
The last term can be rewritten as 
\begin{equation}
\label{eq:Eomega_exp}
E_{\Omega,\mathrm{coh}}(\Delta)=A_0\times \left[
(1-\Omega\eta^2)\cdot\frac{\Gamma/2-i\Delta}{\Delta^2+\Gamma^2/4}
+\Omega\eta^2\cdot\frac{-\Gamma/2-i\Delta}{\Delta^2+\Gamma^2/4}
\right]\quad .
\end{equation}
The first term of this sum is again the spectrum of the incident
increasing exponential pulse but weighted with a proportionality
factor $(1-\Omega\eta^2)$.
The second term which scales with $\Omega\eta^2$ is also the spectrum
of an exponential pulse, but for one with a decaying envelope that
starts at $t=0$. 
In other words, the spectral components of the latter term are the
phase conjugated versions of the incident components.
Although we have assumed that the scattering is completely elastic,
i.e. there is no upper state population, we interpret this
exponentially decaying fraction as absorbed and spontaneously
re-emitted photons. 

The spectral amplitudes in Eqns. (\ref{eq:Ecompl_exp}) and
(\ref{eq:Einc_exp}) correspond to a field envelop that is
increasing with $\exp(\Gamma/(2t))$ for $t<0$ and decreasing with
$\exp(-\Gamma/(2t))$ for $t>0$. 
This is obvious from the expansion
$\Gamma=(\Gamma/2+i\Delta)+(\Gamma/2-i\Delta)$ in the numerator of
these equations, yielding the sum of the spectra corresponding to the
above temporal pulse shapes.
The interpretation of this spectrum is straightforward:
As long as the incident pulse is nonzero,
$E_{\mathrm{sc},\overline\Omega}$ is given by the elastically
scattered incident wave with increasing exponential envelop.
For $t>0$ there is no incident field amplitude, hence the exponential
decay is again interpreted to 'mimic' spontaneous emission.
The same interpretation applies to $E_{\Omega,\mathrm{incoh}}$.
In a recent experiment~\cite{aljunid2013} the light scattered by a
single atom into the backward solid angle $(1-\Omega)$ has been
measured for focusing from $\Omega=0.11$.
The temporal evolution obtained for exponentially increasing coherent
state pulses containing three photons on average resembles the double
sided exponential corresponding to the spectrum in
Eq.~(\ref{eq:Ecompl_exp}).

Summing up the power of all exponentially decaying contributions of
$E_{\Omega,\mathrm{coh}}$, $E_{\Omega,\mathrm{incoh}}$ and
  $E_{\mathrm{sc},\overline\Omega}$ and normalizing to the total power 
  shows that the power fraction in the temporally decaying signal is given
  by $\Omega\eta^2$. 
This suggests to interpret $\Omega\eta^2$ as the absorption
probability of an exponentially increasing pulse that is temporally
'mode-matched' to the atomic transition, which is in accordance with
the findings of Ref.~\cite{golla2012}.

In the limit of focusing from full solid angle, $\Omega=1$,
$E_{\mathrm{sc},\overline\Omega}$ naturally vanishes.
If then also the spatial mode matching becomes perfect, $\eta=1$, also 
$E_{\Omega,\mathrm{incoh}}$ and the first term of
$E_{\Omega,\mathrm{coh}}$ become zero. 
In other words, the remaining spectral field components are completely
given by the spectrum of an exponentially decaying field. 
This is the same result one would obtain from a fully quantized
treatment of the absorption of a single photon by a single atom, if
the incident single photon is the time reversed version of a
spontaneously emitted photon \cite{stobinska2009}.  
We note that the results obtained here are also analogous to the ones 
obtained for the response of an empty Fabry-Perot
resonator~\cite{heugel2010}, which constitutes a fully classical
problem.

\section{Discussion}

By using a fully elastic treatment in describing the scattering of
light by single atoms, we have derived the electric field spectrum
arising from scattering an increasing exponential pulse.
The obtained results suggest to interpret the response as a field
arising via spontaneous emission, which only occurs after absorption
of a photon.
Recent experiments~\cite{aljunid2013} confirm the expected validity of
our framework in the regime of low average photon numbers.
Nevertheless, one has to take into account that in the reported
experiment roughly 11\% of the solid angle was used for focusing.
This means that also the coupling efficiency is limited by this
value~\cite{golla2012,leuchs2013o}.
Therefore, it is not surprising that qualitative agreement with the
model presented here is observed, since approximately only every ninth
photon interacts with the atom.
It will therefore be interesting to compare our theory to the
experiments prepared in Refs.~\cite{golla2012,maiwald2012}, where the
coupling will occur from almost full solid angle and  with large expected
mode overlaps~\cite{golla2012}.
Using the experimental parameters, the relative magnitudes of the
exponentially increasing and decreasing fractions should resemble the
achieved absorption efficiency.
Moreover, increasing the average photon number of the coherent
state pulses used in the experiment will identify the boundary of 
validity of the fully elastic model presented here.

%\bibliographystyle{apsrev}
%\bibliography{/home/markus/Dokumente/bibtex/optik}

\begin{thebibliography}{22}
\expandafter\ifx\csname natexlab\endcsname\relax\def\natexlab#1{#1}\fi
\expandafter\ifx\csname bibnamefont\endcsname\relax
  \def\bibnamefont#1{#1}\fi
\expandafter\ifx\csname bibfnamefont\endcsname\relax
  \def\bibfnamefont#1{#1}\fi
\expandafter\ifx\csname citenamefont\endcsname\relax
  \def\citenamefont#1{#1}\fi
\expandafter\ifx\csname url\endcsname\relax
  \def\url#1{\texttt{#1}}\fi
\expandafter\ifx\csname urlprefix\endcsname\relax\def\urlprefix{URL }\fi
\providecommand{\bibinfo}[2]{#2}
\providecommand{\eprint}[2][]{\url{#2}}

\bibitem[{\citenamefont{Vamivakas et~al.}(2007)\citenamefont{Vamivakas,
  Atat\"ure, Dreiser, Yilmaz, Badolato, Swan, Goldberg, Imamoglu, and
  \"Unl\"u}}]{vamivakas2007}
\bibinfo{author}{\bibfnamefont{A.~N.} \bibnamefont{Vamivakas}},
  \bibinfo{author}{\bibfnamefont{M.}~\bibnamefont{Atat\"ure}},
  \bibinfo{author}{\bibfnamefont{J.}~\bibnamefont{Dreiser}},
  \bibinfo{author}{\bibfnamefont{S.~T.} \bibnamefont{Yilmaz}},
  \bibinfo{author}{\bibfnamefont{A.}~\bibnamefont{Badolato}},
  \bibinfo{author}{\bibfnamefont{A.~K.} \bibnamefont{Swan}},
  \bibinfo{author}{\bibfnamefont{B.~B.} \bibnamefont{Goldberg}},
  \bibinfo{author}{\bibfnamefont{A.}~\bibnamefont{Imamoglu}}, \bibnamefont{and}
  \bibinfo{author}{\bibfnamefont{M.~S.} \bibnamefont{\"Unl\"u}},
  \bibinfo{journal}{Nano Letters} \textbf{\bibinfo{volume}{7}},
  \bibinfo{pages}{2892} (\bibinfo{year}{2007}).

\bibitem[{\citenamefont{Wrigge et~al.}(2008)\citenamefont{Wrigge, Gerhardt,
  Hwang, Zumofen, and Sandoghdar}}]{wrigge2008}
\bibinfo{author}{\bibfnamefont{G.}~\bibnamefont{Wrigge}},
  \bibinfo{author}{\bibfnamefont{I.}~\bibnamefont{Gerhardt}},
  \bibinfo{author}{\bibfnamefont{J.}~\bibnamefont{Hwang}},
  \bibinfo{author}{\bibfnamefont{G.}~\bibnamefont{Zumofen}}, \bibnamefont{and}
  \bibinfo{author}{\bibfnamefont{V.}~\bibnamefont{Sandoghdar}},
  \bibinfo{journal}{Nature Physics} \textbf{\bibinfo{volume}{4}},
  \bibinfo{pages}{60} (\bibinfo{year}{2008}).

\bibitem[{\citenamefont{Tey et~al.}(2008)\citenamefont{Tey, Chen, Aljunid,
  Chng, Huber, Maslennikov, and Kurtsiefer}}]{tey2008-np}
\bibinfo{author}{\bibfnamefont{M.~K.} \bibnamefont{Tey}},
  \bibinfo{author}{\bibfnamefont{Z.}~\bibnamefont{Chen}},
  \bibinfo{author}{\bibfnamefont{S.~A.} \bibnamefont{Aljunid}},
  \bibinfo{author}{\bibfnamefont{B.}~\bibnamefont{Chng}},
  \bibinfo{author}{\bibfnamefont{F.}~\bibnamefont{Huber}},
  \bibinfo{author}{\bibfnamefont{G.}~\bibnamefont{Maslennikov}},
  \bibnamefont{and}
  \bibinfo{author}{\bibfnamefont{C.}~\bibnamefont{Kurtsiefer}},
  \bibinfo{journal}{Nature Physics} \textbf{\bibinfo{volume}{4}},
  \bibinfo{pages}{924} (\bibinfo{year}{2008}).

\bibitem[{\citenamefont{Aljunid et~al.}(2009)\citenamefont{Aljunid, Tey, Chng,
  Liew, Maslennikov, Scarani, and Kurtsiefer}}]{aljunid2009}
\bibinfo{author}{\bibfnamefont{S.~A.} \bibnamefont{Aljunid}},
  \bibinfo{author}{\bibfnamefont{M.~K.} \bibnamefont{Tey}},
  \bibinfo{author}{\bibfnamefont{B.}~\bibnamefont{Chng}},
  \bibinfo{author}{\bibfnamefont{T.}~\bibnamefont{Liew}},
  \bibinfo{author}{\bibfnamefont{G.}~\bibnamefont{Maslennikov}},
  \bibinfo{author}{\bibfnamefont{V.}~\bibnamefont{Scarani}}, \bibnamefont{and}
  \bibinfo{author}{\bibfnamefont{C.}~\bibnamefont{Kurtsiefer}},
  \bibinfo{journal}{Physical Review Letters} \textbf{\bibinfo{volume}{103}},
  \bibinfo{eid}{153601} (\bibinfo{year}{2009}).

\bibitem[{\citenamefont{Slodi\ifmmode~\check{c}\else \v{c}\fi{}ka
  et~al.}(2010)\citenamefont{Slodi\ifmmode~\check{c}\else \v{c}\fi{}ka,
  H\'etet, Gerber, Hennrich, and Blatt}}]{slodicka2010}
\bibinfo{author}{\bibfnamefont{L.}~\bibnamefont{Slodi\ifmmode~\check{c}\else
  \v{c}\fi{}ka}}, \bibinfo{author}{\bibfnamefont{G.}~\bibnamefont{H\'etet}},
  \bibinfo{author}{\bibfnamefont{S.}~\bibnamefont{Gerber}},
  \bibinfo{author}{\bibfnamefont{M.}~\bibnamefont{Hennrich}}, \bibnamefont{and}
  \bibinfo{author}{\bibfnamefont{R.}~\bibnamefont{Blatt}},
  \bibinfo{journal}{Phys. Rev. Lett.} \textbf{\bibinfo{volume}{105}},
  \bibinfo{pages}{153604} (\bibinfo{year}{2010}).

\bibitem[{\citenamefont{Pototschnig et~al.}(2011)\citenamefont{Pototschnig,
  Chassagneux, Hwang, Zumofen, Renn, and Sandoghdar}}]{pototschnig2011}
\bibinfo{author}{\bibfnamefont{M.}~\bibnamefont{Pototschnig}},
  \bibinfo{author}{\bibfnamefont{Y.}~\bibnamefont{Chassagneux}},
  \bibinfo{author}{\bibfnamefont{J.}~\bibnamefont{Hwang}},
  \bibinfo{author}{\bibfnamefont{G.}~\bibnamefont{Zumofen}},
  \bibinfo{author}{\bibfnamefont{A.}~\bibnamefont{Renn}}, \bibnamefont{and}
  \bibinfo{author}{\bibfnamefont{V.}~\bibnamefont{Sandoghdar}},
  \bibinfo{journal}{Phys. Rev. Lett.} \textbf{\bibinfo{volume}{107}},
  \bibinfo{pages}{063001} (\bibinfo{year}{2011}).

\bibitem[{\citenamefont{Rezus et~al.}(2012)\citenamefont{Rezus, Walt, Lettow,
  Renn, Zumofen, G\"otzinger, and Sandoghdar}}]{rezus2012}
\bibinfo{author}{\bibfnamefont{Y.~L.~A.} \bibnamefont{Rezus}},
  \bibinfo{author}{\bibfnamefont{S.~G.} \bibnamefont{Walt}},
  \bibinfo{author}{\bibfnamefont{R.}~\bibnamefont{Lettow}},
  \bibinfo{author}{\bibfnamefont{A.}~\bibnamefont{Renn}},
  \bibinfo{author}{\bibfnamefont{G.}~\bibnamefont{Zumofen}},
  \bibinfo{author}{\bibfnamefont{S.}~\bibnamefont{G\"otzinger}},
  \bibnamefont{and}
  \bibinfo{author}{\bibfnamefont{V.}~\bibnamefont{Sandoghdar}},
  \bibinfo{journal}{Phys. Rev. Lett.} \textbf{\bibinfo{volume}{108}},
  \bibinfo{pages}{093601} (\bibinfo{year}{2012}).

\bibitem[{\citenamefont{Aljunid et~al.}(2013)\citenamefont{Aljunid,
  Maslennikov, Wang, Lan, Scarani, and Kurtsiefer}}]{aljunid2013}
\bibinfo{author}{\bibfnamefont{S.~A.} \bibnamefont{Aljunid}},
  \bibinfo{author}{\bibfnamefont{G.}~\bibnamefont{Maslennikov}},
  \bibinfo{author}{\bibfnamefont{Y.}~\bibnamefont{Wang}},
  \bibinfo{author}{\bibfnamefont{D.~H.} \bibnamefont{Lan}},
  \bibinfo{author}{\bibfnamefont{V.}~\bibnamefont{Scarani}}, \bibnamefont{and}
  \bibinfo{author}{\bibfnamefont{C.}~\bibnamefont{Kurtsiefer}},
  \bibinfo{journal}{arXiv:1304.3761}  (\bibinfo{year}{2013}).

\bibitem[{\citenamefont{Stobinska et~al.}(2009)\citenamefont{Stobinska, Alber,
  and Leuchs}}]{stobinska2009}
\bibinfo{author}{\bibfnamefont{M.}~\bibnamefont{Stobinska}},
  \bibinfo{author}{\bibfnamefont{G.}~\bibnamefont{Alber}}, \bibnamefont{and}
  \bibinfo{author}{\bibfnamefont{G.}~\bibnamefont{Leuchs}},
  \bibinfo{journal}{EPL} \textbf{\bibinfo{volume}{86}}, \bibinfo{pages}{14007}
  (\bibinfo{year}{2009}).

\bibitem[{\citenamefont{Heugel et~al.}(2010)\citenamefont{Heugel, Villar,
  Sondermann, Peschel, and Leuchs}}]{heugel2010}
\bibinfo{author}{\bibfnamefont{S.}~\bibnamefont{Heugel}},
  \bibinfo{author}{\bibfnamefont{A.~S.} \bibnamefont{Villar}},
  \bibinfo{author}{\bibfnamefont{M.}~\bibnamefont{Sondermann}},
  \bibinfo{author}{\bibfnamefont{U.}~\bibnamefont{Peschel}}, \bibnamefont{and}
  \bibinfo{author}{\bibfnamefont{G.}~\bibnamefont{Leuchs}},
  \bibinfo{journal}{Laser Physics} \textbf{\bibinfo{volume}{20}},
  \bibinfo{pages}{100} (\bibinfo{year}{2010}).

\bibitem[{\citenamefont{Wang et~al.}(2011)\citenamefont{Wang, Min\'a\v{r},
  Sheridan, and Scarani}}]{wang2011}
\bibinfo{author}{\bibfnamefont{Y.}~\bibnamefont{Wang}},
  \bibinfo{author}{\bibfnamefont{J.}~\bibnamefont{Min\'a\v{r}}},
  \bibinfo{author}{\bibfnamefont{L.}~\bibnamefont{Sheridan}}, \bibnamefont{and}
  \bibinfo{author}{\bibfnamefont{V.}~\bibnamefont{Scarani}},
  \bibinfo{journal}{Phys. Rev. A} \textbf{\bibinfo{volume}{83}},
  \bibinfo{pages}{063842} (\bibinfo{year}{2011}).

\bibitem[{\citenamefont{Leuchs and Sondermann}(2013)}]{leuchs2013o}
\bibinfo{author}{\bibfnamefont{G.}~\bibnamefont{Leuchs}} \bibnamefont{and}
  \bibinfo{author}{\bibfnamefont{M.}~\bibnamefont{Sondermann}},
  \bibinfo{journal}{Journal of Modern Optics} \textbf{\bibinfo{volume}{60}},
  \bibinfo{pages}{36} (\bibinfo{year}{2013}).

\bibitem[{\citenamefont{Golla et~al.}(2012)\citenamefont{Golla, Chalopin,
  Bader, Harder, Mantel, Maiwald, Lindlein, Sondermann, and
  Leuchs}}]{golla2012}
\bibinfo{author}{\bibfnamefont{A.}~\bibnamefont{Golla}},
  \bibinfo{author}{\bibfnamefont{B.}~\bibnamefont{Chalopin}},
  \bibinfo{author}{\bibfnamefont{M.}~\bibnamefont{Bader}},
  \bibinfo{author}{\bibfnamefont{I.}~\bibnamefont{Harder}},
  \bibinfo{author}{\bibfnamefont{K.}~\bibnamefont{Mantel}},
  \bibinfo{author}{\bibfnamefont{R.}~\bibnamefont{Maiwald}},
  \bibinfo{author}{\bibfnamefont{N.}~\bibnamefont{Lindlein}},
  \bibinfo{author}{\bibfnamefont{M.}~\bibnamefont{Sondermann}},
  \bibnamefont{and} \bibinfo{author}{\bibfnamefont{G.}~\bibnamefont{Leuchs}},
  \bibinfo{journal}{Eur. Phys. J. D} \textbf{\bibinfo{volume}{66}},
  \bibinfo{pages}{190} (\bibinfo{year}{2012}).

\bibitem[{\citenamefont{Sondermann et~al.}(2007)\citenamefont{Sondermann,
  Maiwald, Konermann, Lindlein, Peschel, and Leuchs}}]{sondermann2007}
\bibinfo{author}{\bibfnamefont{M.}~\bibnamefont{Sondermann}},
  \bibinfo{author}{\bibfnamefont{R.}~\bibnamefont{Maiwald}},
  \bibinfo{author}{\bibfnamefont{H.}~\bibnamefont{Konermann}},
  \bibinfo{author}{\bibfnamefont{N.}~\bibnamefont{Lindlein}},
  \bibinfo{author}{\bibfnamefont{U.}~\bibnamefont{Peschel}}, \bibnamefont{and}
  \bibinfo{author}{\bibfnamefont{G.}~\bibnamefont{Leuchs}},
  \bibinfo{journal}{Appl. Phys. B} \textbf{\bibinfo{volume}{89}},
  \bibinfo{pages}{489} (\bibinfo{year}{2007}).

\bibitem[{\citenamefont{Sondermann and Leuchs}(2013)}]{sondermann2013p}
\bibinfo{author}{\bibfnamefont{M.}~\bibnamefont{Sondermann}} \bibnamefont{and}
  \bibinfo{author}{\bibfnamefont{G.}~\bibnamefont{Leuchs}},
  \bibinfo{journal}{arXiv:1306.2804 [quant-ph]}  (\bibinfo{year}{2013}).

\bibitem[{\citenamefont{Sondermann et~al.}(2008)\citenamefont{Sondermann,
  Lindlein, and Leuchs}}]{sondermann2008}
\bibinfo{author}{\bibfnamefont{M.}~\bibnamefont{Sondermann}},
  \bibinfo{author}{\bibfnamefont{N.}~\bibnamefont{Lindlein}}, \bibnamefont{and}
  \bibinfo{author}{\bibfnamefont{G.}~\bibnamefont{Leuchs}},
  \bibinfo{journal}{arXiv:0811.2098 [physics.optics]}  (\bibinfo{year}{2008}).

\bibitem[{\citenamefont{Zumofen et~al.}(2008)\citenamefont{Zumofen, Mojarad,
  Sandoghdar, and Agio}}]{zumofen2008}
\bibinfo{author}{\bibfnamefont{G.}~\bibnamefont{Zumofen}},
  \bibinfo{author}{\bibfnamefont{N.~M.} \bibnamefont{Mojarad}},
  \bibinfo{author}{\bibfnamefont{V.}~\bibnamefont{Sandoghdar}},
  \bibnamefont{and} \bibinfo{author}{\bibfnamefont{M.}~\bibnamefont{Agio}},
  \bibinfo{journal}{Phys. Rev. Lett.} \textbf{\bibinfo{volume}{101}},
  \bibinfo{pages}{180404} (\bibinfo{year}{2008}).

\bibitem[{\citenamefont{Tey et~al.}(2009)\citenamefont{Tey, Maslennikov, Liew,
  Aljunid, Huber, Chng, Chen, Scarani, and Kurtsiefer}}]{tey2009}
\bibinfo{author}{\bibfnamefont{M.~K.} \bibnamefont{Tey}},
  \bibinfo{author}{\bibfnamefont{G.}~\bibnamefont{Maslennikov}},
  \bibinfo{author}{\bibfnamefont{T.~C.~H.} \bibnamefont{Liew}},
  \bibinfo{author}{\bibfnamefont{S.~A.} \bibnamefont{Aljunid}},
  \bibinfo{author}{\bibfnamefont{F.}~\bibnamefont{Huber}},
  \bibinfo{author}{\bibfnamefont{B.}~\bibnamefont{Chng}},
  \bibinfo{author}{\bibfnamefont{Z.}~\bibnamefont{Chen}},
  \bibinfo{author}{\bibfnamefont{V.}~\bibnamefont{Scarani}}, \bibnamefont{and}
  \bibinfo{author}{\bibfnamefont{C.}~\bibnamefont{Kurtsiefer}},
  \bibinfo{journal}{New Journal of Physics} \textbf{\bibinfo{volume}{11}},
  \bibinfo{pages}{043011} (\bibinfo{year}{2009}).

\bibitem[{\citenamefont{Zumofen et~al.}(2009)\citenamefont{Zumofen, Mojarad,
  and Agio}}]{zumofen2009}
\bibinfo{author}{\bibfnamefont{G.}~\bibnamefont{Zumofen}},
  \bibinfo{author}{\bibfnamefont{N.~M.} \bibnamefont{Mojarad}},
  \bibnamefont{and} \bibinfo{author}{\bibfnamefont{M.}~\bibnamefont{Agio}},
  \bibinfo{journal}{Nuovo Cimento C} \textbf{\bibinfo{volume}{31}},
  \bibinfo{pages}{475} (\bibinfo{year}{2009}).

\bibitem[{\citenamefont{Jechow et~al.}(2013)\citenamefont{Jechow, Norton,
  H\"andel, Bl\ifmmode~\bar{u}\else \={u}\fi{}ms, Streed, and
  Kielpinski}}]{jechow2013}
\bibinfo{author}{\bibfnamefont{A.}~\bibnamefont{Jechow}},
  \bibinfo{author}{\bibfnamefont{B.~G.} \bibnamefont{Norton}},
  \bibinfo{author}{\bibfnamefont{S.}~\bibnamefont{H\"andel}},
  \bibinfo{author}{\bibfnamefont{V.}~\bibnamefont{Bl\ifmmode~\bar{u}\else
  \={u}\fi{}ms}}, \bibinfo{author}{\bibfnamefont{E.~W.} \bibnamefont{Streed}},
  \bibnamefont{and}
  \bibinfo{author}{\bibfnamefont{D.}~\bibnamefont{Kielpinski}},
  \bibinfo{journal}{Phys. Rev. Lett.} \textbf{\bibinfo{volume}{110}},
  \bibinfo{pages}{113605} (\bibinfo{year}{2013}).

\bibitem[{\citenamefont{Tyc}(2012)}]{tyc2012}
\bibinfo{author}{\bibfnamefont{T.}~\bibnamefont{Tyc}}, \bibinfo{journal}{Opt.
  Lett.} \textbf{\bibinfo{volume}{37}}, \bibinfo{pages}{924}
  (\bibinfo{year}{2012}).

\bibitem[{\citenamefont{Maiwald et~al.}(2012)\citenamefont{Maiwald, Golla,
  Fischer, Bader, Heugel, Chalopin, Sondermann, and Leuchs}}]{maiwald2012}
\bibinfo{author}{\bibfnamefont{R.}~\bibnamefont{Maiwald}},
  \bibinfo{author}{\bibfnamefont{A.}~\bibnamefont{Golla}},
  \bibinfo{author}{\bibfnamefont{M.}~\bibnamefont{Fischer}},
  \bibinfo{author}{\bibfnamefont{M.}~\bibnamefont{Bader}},
  \bibinfo{author}{\bibfnamefont{S.}~\bibnamefont{Heugel}},
  \bibinfo{author}{\bibfnamefont{B.}~\bibnamefont{Chalopin}},
  \bibinfo{author}{\bibfnamefont{M.}~\bibnamefont{Sondermann}},
  \bibnamefont{and} \bibinfo{author}{\bibfnamefont{G.}~\bibnamefont{Leuchs}},
  \bibinfo{journal}{Phys. Rev. A} \textbf{\bibinfo{volume}{86}},
  \bibinfo{pages}{043431} (\bibinfo{year}{2012}).

\end{thebibliography}

\end{document}